\documentclass[showpacs,aps,twocolumn]{revtex4}
\usepackage{graphicx}
\usepackage{amsfonts}
\usepackage{amsmath}
\usepackage{amssymb}
\usepackage{multirow}
\usepackage{braket}
\usepackage[usenames,dvipsnames]{color}
\usepackage[english]{babel}
\usepackage[T1]{fontenc}

\begin{document}
\bibliographystyle{apsrev}

\title{Memory-dependent noise-induced resonance and diffusion in non-markovian systems}

\author{S.~S.~Melnyk}
\affiliation{O. Ya. Usikov Institute for Radiophysics and Electronics NASU, 61085 Kharkiv, Ukraine}

\author{O.~V.~Usatenko}
\affiliation{O. Ya. Usikov Institute for Radiophysics and Electronics NASU, 61085 Kharkiv, Ukraine}

\author{V.~A.~Yampol'skii}
\affiliation{O. Ya. Usikov Institute for Radiophysics and
Electronics NASU, 61085 Kharkov, Ukraine} \affiliation{V. N. Karazin
Kharkov National University, 61077 Kharkov, Ukraine}

\begin{abstract}
We study the random processes with non-local memory and obtain new
solutions of the Mori-Zwanzig equation describing non-markovian
systems. We analyze the system dynamics depending on the amplitudes
$\nu$ and $\mu_0$ of the local and non-local memory and pay
attention to the line in the ($\nu$, $\mu_0$)-plane separating the
regions with asymptotically stationary and non-stationary behavior.
We obtain general equations for such boundaries and consider them
for three examples of the non-local memory functions. We show that
there exist two types of the boundaries with fundamentally different
system dynamics. On the boundaries of the first type, the diffusion
with memory takes place, whereas on borderlines of the second type,
the phenomenon of noise-induced resonance can be observed. A
distinctive feature of noise-induced resonance in the systems under
consideration is that it occurs in the absence of an external
regular periodic force. It takes place due to the presence  of
frequencies in the noise spectrum, which are close to the
self-frequency of the system. We analyze also the variance of the
process and compare its behavior for regions of asymptotic
stationarity and non-stationarity, as well as for diffusive and
noise-induced-resonance borderlines between them.

\end{abstract}


\maketitle

\section{Introduction}

The Markov processes are the simplest and the most popular models
for describing the random phenomena (see, e.g.,
Refs.~\cite{zab,reb,fer,nic,Uhlen,Kampen,gar,Horsthemke}). A lot of
systems in the real world are more complex than the markovian ones,
they have non-markovian character of the memory (see, e.g.,
Refs.~\cite{mok,bre,Sarracino_2010,sie,ros,sta,Caceres,Kampen98}).
Therefore, it is necessary to go beyond the simple markovian model.
In recent years, a lot of attention has been paid to studying the
non-Markov processes, in particular, due to their role in
decoherence phenomena in open quantum systems (see, e.g.,
Refs.~\cite{lam,bre,kan}). Namely, non-markovianity can serve as a
source for suppressing the exponential decay of coherence in the
interaction of a quantum system with a classical thermal
bath~\cite{bel,chi,byl}.

In formulation of what is the Markov process, very important role is
played by its exponential correlation function. As was shown in
Refs.~\cite{HanggiStatPhys,Nakazawa}, the replacement of the
exponential correlation function by another one leads to the
non-stationarity of the process. A particular class of strongly
non-markovian stochastic processes with long-range correlated noise
appearing in the corresponding stochastic differential equation
(SDE) was studied in Refs.~\cite{Wang,Goychuk1}.
McCauley~\cite{McCauley} considered the non-stationary non-markovian
processes with 1-state memory where the SDE takes into account the
value of random variable $V$ at fixed temporal point $t_0$ in the
past.

The difficulties arising in attempts to introduce a correlation
function different from exponential are closely connected with two
facts: a desire to determine the conditional probability
distribution function (CPDF) for arbitrary time laps $\tau$ from the
last known value of random variable and to determine a group chain
rule for the CPDF. To overcome these difficulties, we have
introduced in Ref.~\cite{MYaU} integral memory term depending on
the past of the process into the SDE and the transition probability
function. Thus, we refused to deal with the CPDF  for arbitrary
value $\tau$ and considered the case of infinitesimal $\tau = dt
\rightarrow 0$ only.

Introduction of the integral memory term results in transformation of the
SDE into the stochastic integro-differential equation (SIDE),
\begin{eqnarray} \label{SIDE}
dV(t)&=&-\nu V(t) dt \\[6pt]
 & - &\int_{0} ^{\infty} \mu(t')
V(t-t')dt'dt +\sigma \, dW(t).\nonumber
\end{eqnarray}
Here $dW(t)$ is the standard white noise, i.e., $W(t)$ is the
continuous centered Wiener process with independent increments with
variance $ \langle (W(t+\tau)-W(t))^2\rangle = |\tau|$, or,
equivalently, $W(t)= \int dW(t) \Rightarrow $ $\langle
dW(t)dW(t')\rangle =\delta(t-t')dtdt'$, the symbol $\langle ...
\rangle$ denotes a statistical ensemble averaging. The term $-\nu
V(t) dt$ in Eq.~\eqref{SIDE} describes a local-memory one-point
feature of the process. The positive value of the constant $\nu$
provides an anti-persistent character of the process with attraction
of $V(t)$ to the point $V=0$. If we omit the memory term $\mu(t')$
in Eq.~\eqref{SIDE}, then we obtain the well known equation for the
Ornstein-Uhlenbeck process, which simulates the Brownian motion of a
microscopic particle in a liquid viscous suspension subjected to a
random force with intensity $\sigma$. Equation~\eqref{SIDE} is often
named as the Mori-Zwanzig
one~\cite{Mori1965,Zwanzig1960,Zwanzig2001}, or the
external-regular-force-independent generalized Langevin
equation~\cite{Goychuk1}. The Mori-Zwanzig equation~\eqref{SIDE}
finds numerous applications (see, e.g.,
Ref.~\cite{Vrugt_MoriZwanzig} and references therein).

Such generalization of SDE has also been discussed by many
authors~\cite{Adelman,Hanggi_Thomas,Hynes,Wang}. In most cases, the
so-called internal noise was considered, when, according to the
fluctuation-dissipation theorem~\cite{Kubo}, the function $\mu(t)$
is uniquely determined by the correlation function of the stochastic
perturbation $W(t)$. Then the memory kernel $\mu(t)$ describes the
so-called viscoelastic friction~\cite{Goychuk1}. However, in the
case of external noise, the fluctuation and dissipation come from
different sources, i.e., the frictional kernel $\mu(t)$ and the
correlation function of the noise are independent of each other
(see, e.g., Ref.~\cite{Wang}).

In this paper we consider an \emph{arbitrary} memory kernel $\mu(t)$
and a Gaussian \emph{external} noise $W(t)$ \emph{independent} of
$\mu(t)$. In this case Eq.~\eqref{SIDE} could be a good physical
model for the systems where the external noise is much more
intensive than the thermal one.

Our general consideration of the Mory-Zwanzig equation is
accompanied by the model examples of the memory functions. The first
example is the local memory function defined at the time moment
$(t-T)$ remote at the depth $T$ from the instant time moment $t$,
\begin{eqnarray} \label{103}
\mu (t) = \frac{\mu _0}{T} \delta (t-T).
\end{eqnarray}
Here $\delta(.)$ denotes the Dirac delta, $\mu _0$ is the memory
amplitude. To produce the random value of $V(t+dt)$ the system
``uses'' the knowledge about its past in the points $t$ and $t-T$.
This memory function is a good approximation for any process with
a pronounced maximum in the $\mu(t)$ dependence at $t=T$.

The second example is the step-wise memory
function~\cite{UYa,RewUAMM},
\begin{eqnarray} \label{105}
\mu (t) = \frac{\mu _0}{T^2} \theta (T-t),
\end{eqnarray}
where $\theta(.)$ is the Heaviside theta-function. This function is
a good approximation for any process with a pronounced edge
in the $\mu(t)$ dependence at $t=T$.

At last, we show that Eq.~\eqref{SIDE} has an exact analytical
solution for the memory function of the exponential form,
\begin{equation}\label{exp}
 \mu(t) = \frac{\mu_0}{T^2} \exp(-t/T).
\end{equation}

Note that the exponential memory function can be used to describe many real physical phenomena, e.g., the coupling of a massive tracer with the surrounding granular fluid~\cite{Sarracino_2010}. This model describes qualitatively any other processes with smoothly decreasing memory function. Thus, the considered here three examples of memory functions describe qualitatively the most typical kinds of the $\mu(t)$ dependences, regardless of their physical implementations.

The dynamics of the system described by Eq.~\eqref{SIDE} is very
sensitive to the region in which the parameters $ \mu_0 $ and $ \nu
$ are located. In particular, it was shown in our previous work
\cite{MYaU} that the process with the delta-functional memory is
asymptotically stationary not for any values of $ \mu_0 $ and $ \nu
$. It is very interesting and nontrivial that, for example, for $
\nu = 0 $, there are two boundaries of asymptotic stationarity, $
\mu_0 = 0 $ and $ \mu_0 = \mu_{\mathrm{crit}}= 2/\pi $. Approaching
the lower boundary, we observe the ordinary Brownian diffusion.
Approaching the upper boundary, for $ \mu_0 \rightarrow
\mu_{\mathrm{crit}} $, the process goes into the oscillation mode
with a certain fixed frequency of oscillations. The analysis of
Eqs.~\eqref{Cond_Omega_Harmonic_ReIm}, which are presented in the
next Section, shows that similar two boundaries of stationarity
exist for any system with arbitrary memory function $\mu(t)$.

In this paper, we study the system dynamics in various regions of
the parameters $ \mu_0 $ and $ \nu $ with the main focus on the
boundaries of the region of asymptotic stationarity. We show that
there are two types of such boundaries with fundamentally different
system behavior. On the boundaries of the first type, corresponding
to smaller values of $ \mu_0 $, a diffusion with non-local memory
takes place, and we call these borderlines as diffusive. On the
boundaries of the second type, corresponding to larger values of $
\mu_0 $, the phenomenon of noise-induced resonance occurs.

The scope of the paper is as follows. In the next section, we obtain
general expressions for the boundaries of the region of asymptotic
stationarity in the $ ( \nu, \mu_0 )$-plane, and present these
boundaries for the above mentioned three examples of memory
functions.

In Section III, we analyze the behavior of the system  for different
prehistories in various areas in the $ (\nu,\mu_0) $-plain in the
absence of random force. We show that, on the upper borderline of
the asymptotic stationarity region, the variable $ V(t) $ goes
asymptotically into an oscillatory mode with some given frequency.
This means that we deal here with the system with well-defined
frequency of self-oscillations.  On the lower borderline , the
variable $ V $ tends to a constant value at $t\rightarrow\infty$.

Section IV is the main in our paper. Here we show that the switching
on the random force in the Mori-Zwanzig system leads to the
diffusion on the lower boundary of asymptotic stationarity and to
the noise-induced resonance at the upper boundary. A distinctive
feature of the noise-induced resonance in the systems under
consideration is that it occurs in the absence of an external
regular periodic force. It takes place due to the presence  of
frequencies in the noise spectrum, which are close to the
self-frequency of the system. Then we study the variance of the
process and compare its behavior for regions of asymptotic
stationarity and non-stationarity, as well as for diffusion and
noise-induced-resonance boundaries between them.

\section{Boundaries of asymptotic stationarity}

The random process under study is very sensitive to the values of
two memory parameters, $\nu$ and $\mu_0$. In this section, we
analyze the borderlines of region in the $(\nu,\mu_0)$-plain where
the process is asymptotically stationary. In this region, the
two-point correlation function $C(t_1,t_2)$,
\begin{equation}\label{CF}
 C(t_1,t_2)= \langle V(t_1)V(t_2)\rangle- \langle V(t_1) \rangle \langle V(t_2)
 \rangle,
\end{equation}
is asymptotically dependent on the difference $t_2-t_1 \equiv t$
only, i.e., $C(t_1, t_2) \approx
C(t)$ at $t_1,t_2 \rightarrow \infty$:
\begin{equation}\label{C(t)}
 C(t)=\lim_{t'\rightarrow\infty}C(t',t'+t).
\end{equation}
Herein the time difference $t$ can be arbitrary.

As was shown in Ref.~\cite{MYaU}, the correlation function $C(t)$ of
the process is governed by the continuous analog of the Yule-Walker
equation,~\cite{Yule,Walker},
\begin{equation}\label{AM_C_Phi} \frac{d C(t)}{dt} + \nu C(t) +
\int_0^\infty \mu(t') C(t - t') dt' =0, \quad t > 0,
\end{equation}
with the boundary condition,
\begin{equation}\label{nc}
    \frac{dC(t)}{dt}{\Big|_{t=0_+}} = -\frac{\sigma^2}{2}.
\end{equation}
The argument $0_+$ signifies that the derivative is taken at
positive $t$ close to zero. The simple method to obtain
Eq.~\eqref{AM_C_Phi} is presented in Appendix A.

Two equations, \eqref{AM_C_Phi} and \eqref{nc}, represent a very
useful tool for studying the statistical properties of random
processes with non-local memory. These properties are governed by
the constants $\nu$, $\sigma$, and the memory function $\mu(t)$. We
assume that the function $\mu(t)$ has good properties at
$t\rightarrow\infty$. More exactly, we suppose that the function
$\mu(t)$ has either a finite characteristic scale $T$ of decrease,
or it abruptly vanishes at $t>T$, $\mu(t>T)=0$. In this case, the
correlation function can be presented as a sum of exponential terms,
\begin{eqnarray} \label{101}
C(t) = \sum_i C_i \exp \left( -\frac{z_i t}{T} \right),
\end{eqnarray}
for $t \gg T$.

Equation~\eqref{AM_C_Phi} gives the following characteristic
algebraic equation for the complex decrements $z_i$:
\begin{eqnarray} \label{102}
\frac{z}{T}=\nu +  \int_{0} ^{\infty} \mu(t)\exp \left(\frac{z
t}{T}\right) d t.
\end{eqnarray}
Solving it, we find a set of $z_i$ as functions of the parameters
$\nu$ and $\mu_0$. We are interested in the root $z_0$ of
Eq.~\eqref{102} with the lowest real part because specifically this
root defines behavior of the correlation function Eq.~\eqref{101} at
$t \rightarrow \infty$. From Eq.~\eqref{101}, one can see that the
imaginary part of $z_0=\xi_0 + i\zeta_0$ corresponds to the
oscillations of $C(t)$, while the sign of its real part, $\xi_0$,
defines the stationarity properties. The positive $\xi_0$
corresponds to the exponential decrease of the correlation function
$C(t)$, and the negative value of $\xi_0$ does to the exponential
increase.

Thus, to find the borderline of stationary range in the
$(\nu,\mu_0)$-plain, we should solve Eq.~\eqref{102} for the purely
imaginary $z=i\zeta$. In this case Eq.~\eqref{102} gives
\begin{equation}
\begin{cases}\label{Cond_Omega_Harmonic_ReIm}
\nu + \int_0^\infty \mu(t) \cos \left(\dfrac{\zeta t}{T}\right)  \; dt = 0, \\[8pt]
\dfrac{\zeta}{T}- \int_0^\infty \mu(t) \sin \left(\dfrac{\zeta
t}{T}\right) \; dt =0 .
\end{cases}
\end{equation}

Let us apply the set of Eqs.~\eqref{Cond_Omega_Harmonic_ReIm} for
investigating the stationarity borderlines in the frame of the
mentioned above three models of the non-local memory $\mu(t)$.
\subsubsection{Delta-functional memory}
\begin{figure}[h!]
\center\includegraphics[width=0.5\textwidth]{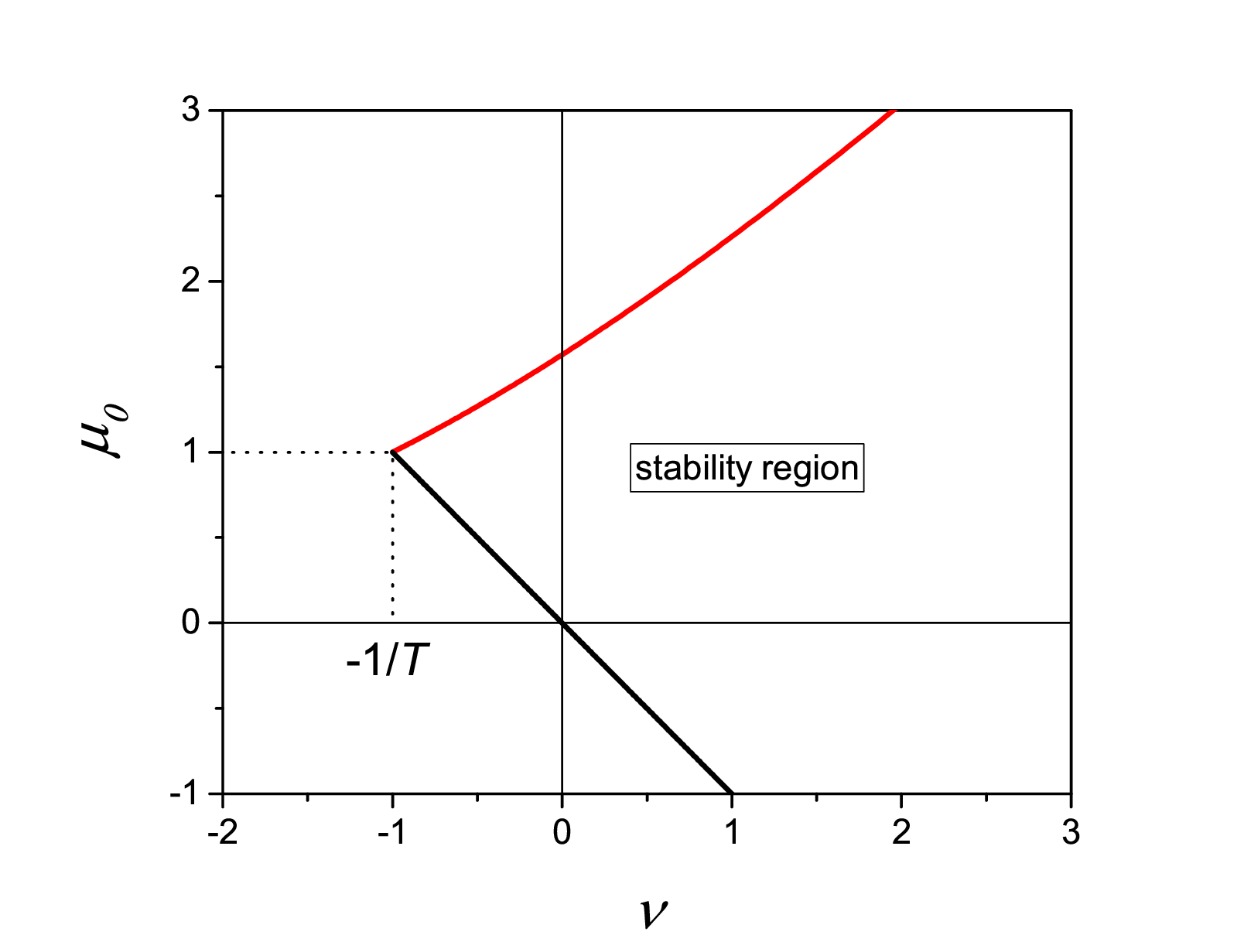} \caption{The
stationarity borderlines for the delta-functional memory $\mu(t)=
(\mu_0/T)\delta(t-T)$ with $T=1$ in the plane $ (\nu, \mu_0) $. The
red solid curve at $\mu_0>1$ corresponds to the oscillatory
borderline, and the black solid straight line does to the diffusive
one.} \label{FigStatArea_delta}
\end{figure}

As the first example, we consider the memory function
$\mu(t)=(\mu_0/T) \delta(t-T)$. Then,
Eq.~\eqref{Cond_Omega_Harmonic_ReIm} transforms into
\begin{equation}\label{Cond_1}
\begin{cases}
\nu T + \mu_0 \cos \zeta = 0, \\[5pt]
\zeta - \mu_0 \sin \zeta = 0.
\end{cases}
\end{equation}
For $ 0 < \zeta < \pi$ this set of equations describes the so called
``oscillatory'' borderline because the corresponding correlation
function $C(t)$, Eq.~\eqref{101}, oscillates without damping when
approaching this borderline. In the case $\zeta \rightarrow 0$, the
$C(t)$ function tends very smoothly to zero without oscillations in
the vicinity of stationarity borderline. Assuming $\zeta = 0$ in
Eq.~\eqref{Cond_1}, we get for this borderline,
\begin{equation}\label{BorderDif}
\nu T + \mu_0 = 0.
\end{equation}
Figure~\ref{FigStatArea_delta} shows the oscillatory (upper red
curve) and diffusive (lower straight black line) stationarity
borderlines.

Note that the general equation, valid for arbitrary memory function,
describing the diffusive borderline, can easily be obtained if we
put $\zeta=0$ in Eqs.~\eqref{Cond_Omega_Harmonic_ReIm},
\begin{equation}\label{nu_mu}
\nu + \int_0^\infty \mu(t) dt = 0.
\end{equation}
If $\int_0^\infty \mu(t) dt\neq 0$,  we can define the amplitude
$\mu_0$ of the memory function as
\begin{equation}\label{def_mu_0}
\mu_0 = T \int_0^\infty \mu(t) dt.
\end{equation}
Then Eq.~\eqref{BorderDif} for the diffusive borderline will be
valid for any memory function.

\subsubsection{Step-wise memory function}

\begin{figure}[h!]
\center\includegraphics[width=0.5\textwidth]{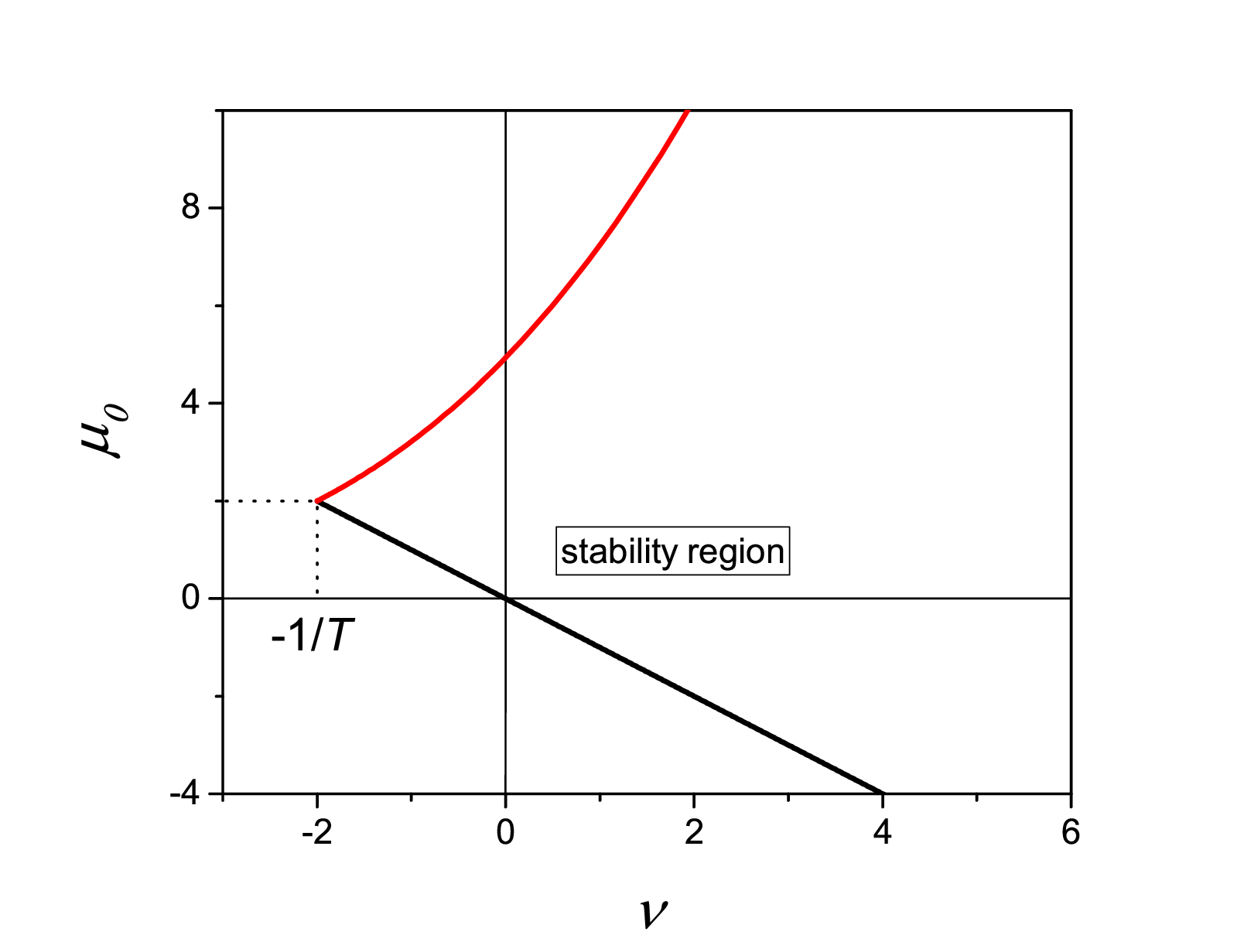} \caption{The
stationarity borderlines for the step-wise memory function
$\mu(t)=(\mu_0/T^2) \theta(T-t)$ with $T=0.5$ in the plane $ (\nu,
\mu_0) $. The upper red solid curve is the oscillatory borderline,
and the lower black solid straight line at $\mu_0< 2$ is the
diffusive one.} \label{Fig_Zones_StepWise}
\end{figure}

As the second example, we consider the step-wise memory function $
\mu(t) = (\mu_0/T^2) \theta(T-t)$. From the same considerations as
above, we obtain the following relations:
\begin{equation}\label{StepRes}
\begin{cases}
\nu = - \dfrac{1}{T}\dfrac{\zeta \sin \zeta}{1-\cos \zeta}, \\[6pt]
\mu_0 = \dfrac{\zeta^2}{1-\cos \zeta}, \qquad 0\leqslant \zeta <2
\pi,
\end{cases}
\end{equation}
for the oscillatory borderline and Eq.~\eqref{BorderDif} for the
diffusive one. These two borderlines are shown in
Fig.~\ref{Fig_Zones_StepWise}.

\subsubsection{Exponential memory function}\label{ExpMem}

\begin{figure}[h!]
\center\includegraphics[width=0.5\textwidth]{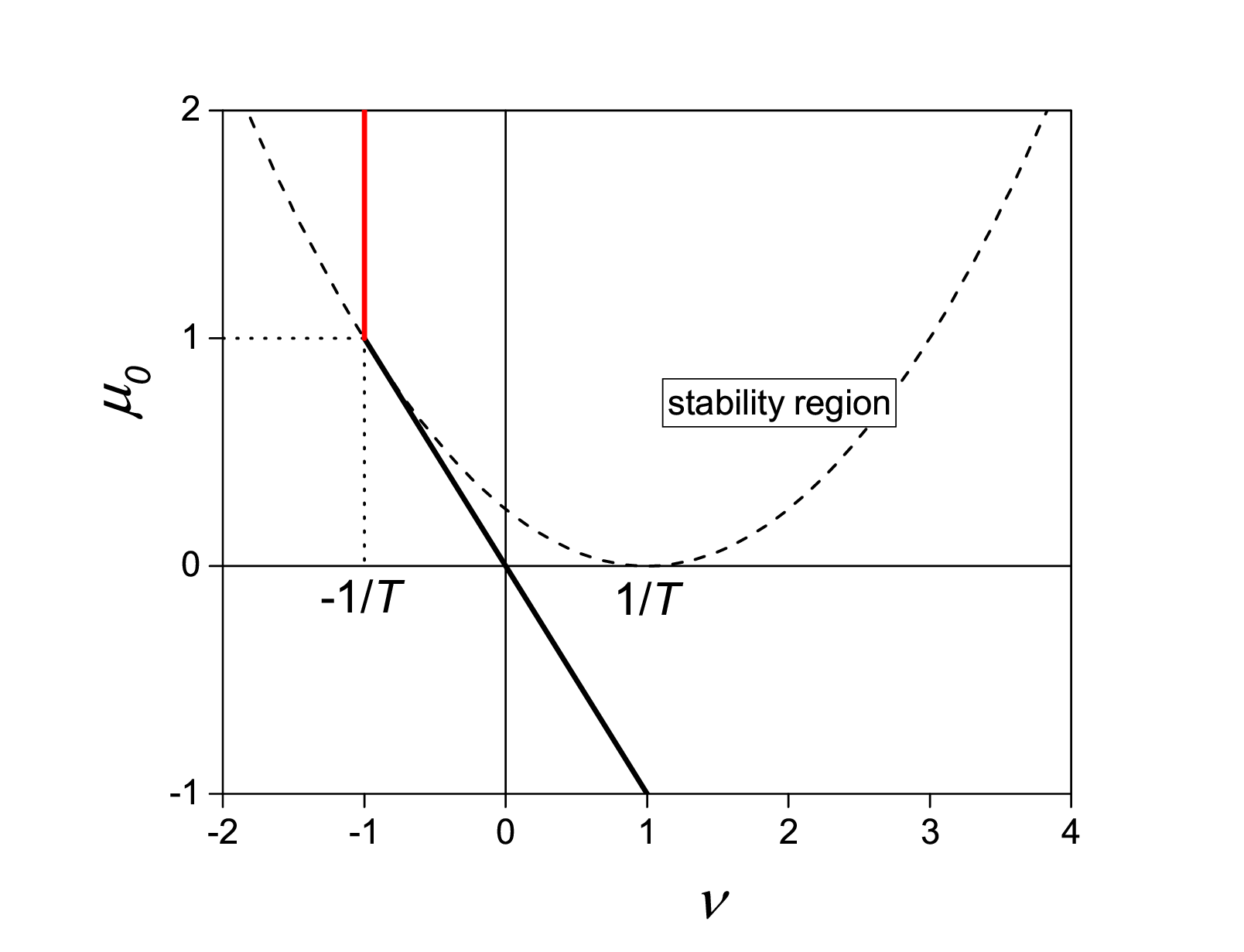} \caption{The
stationarity borderlines for the exponential memory function $ \mu
(t) = (\mu_0/T^2) \exp(-t/T)$ with $T=1$ in the plane $ (\nu, \mu_0)
$. The region of stationarity lies to the right of the solid line,
the region of non-stationarity lies to the left of this line. The
vertical red and oblique black solid lines correspond to the oscillatory
and diffusive borderlines, respectively. A dashed parabola separates the areas where the correlation
function decays exponentially without oscillations (below the
parabola) and with oscillations (above this curve). }
\label{ExpZones}
\end{figure}

As the third example, we consider the exponential memory function
$\mu(t)=(\mu_0/T^2) \exp(-t/T)$ with the positive memory depth $T$.
Then the condition for the diffusive borderline is
Eq.~\eqref{BorderDif}. For the oscillatory borderline we have
\begin{equation}
\begin{cases}
\nu = - \dfrac{1}{T}, \\
\mu_0 = 1+ \zeta^2.
\end{cases}
\end{equation}
These two borderlines are shown in Fig.~\ref{ExpZones}.

Thus, the results obtained in this Section are as follows:

\begin{itemize}

\item The correlation function $C(t)$ of the random process with non-local
memory can be presented as a sum of exponential functions with the
complex decrements/increments $z_i$ defined by Eq.~\eqref{102}.

\item The stationarity of the process is defined by the root
$z_0$ of Eq.~\eqref{102} with the smallest real part. If $\xi_0=\Re
z_0 > 0$, then the function $C(t\rightarrow\infty)$ tends to zero,
and the stochastic process $V(t)$ is stationary. If $\xi_0 < 0$,
then the process $V(t)$ is non-stationary.

\item The condition $\xi_0 = 0$ defines the borderlines between the stationary
and non-stationary regions in the ($\nu$, $\mu_0$)-plain. There
exist two types of borderlines, diffusive and oscillatory ones. The
diffusive borderline corresponds to the case when the imaginary part
of $z_0$ equals zero, $\zeta_0= \Im z_0 = 0$.  This borderline is
described by Eq.~\eqref{nu_mu} (see black solid straight lines in
Figs.~\ref{FigStatArea_delta}, \ref{Fig_Zones_StepWise}, and
\ref{ExpZones} for the examples considered above). The oscillatory
borderline corresponds to $\zeta_0 \neq 0$ and is described by
Eq.~\eqref{Cond_Omega_Harmonic_ReIm} (see red solid curves in
Figs.~\ref{FigStatArea_delta}, \ref{Fig_Zones_StepWise}, and
\ref{ExpZones} for the examples considered above).

\item When approaching the diffusive borderline, the random process goes
to the diffusion with memory and the decrement of $C(t)$ tends to
zero. Approaching the oscillatory borderline, the correlation
function goes into the oscillation mode with a certain frequency of
oscillations.

\item The conditions of stationarity for the process are independent of the
random-force intensity $\sigma$.

\end{itemize}

\section{Movement in the absence of random force}\label{EMF1}
In this Section, we analyze the system dynamics for different
 prehistories (i.e., for different $V(t)$ dependences at
$t\leqslant0$) in various areas of the $(\nu,\mu_0)$-plain in the
absence of random force. We show that,  on the diffusive borderline,
the variable $V(t\rightarrow \infty)$  reaches the constant value.
On the oscillatory borderline, the variable $V(t\rightarrow \infty)$
goes into oscillatory mode with some given frequency. This means
that in the latter case we deal with the specific linear oscillatory
system.

\subsection{Exact fundamental solution}\label{EMF2}
The exact fundamental solution of deterministic (without external
random force $dW(t)$) version of Eq.~\eqref{SIDE},
\begin{eqnarray} \label{SIDEdet}
\frac{dV(t)}{dt}=-\nu V(t)    - \int_{0} ^{\infty} \mu(t')
V(t-t')dt',
\end{eqnarray}
with the fundamental prehistory,
\begin{equation}\label{fundam}
V(t\leqslant 0)=
\begin{cases}
0, &t<0,\\
1,&t= 0,
\end{cases}
\end{equation}
can be found by the method of Laplace transformation (see, e.g.,
Ref.~\cite{Wang}). Denoting this solution by $h(t)$ and performing
the Laplace transformation of Eq.~\eqref{SIDEdet}, we obtain the
image $\tilde{h}(p)$ in the form,
\begin{equation}\label{h_mu_Laplass}
\tilde{h}(p)=\int _0^\infty h(t )\exp(-p\,t) dt =\frac{1}{p + \nu
+\tilde{\mu}(p)},
\end{equation}
where $\tilde{\mu}(p)$ is the Laplace image of the memory function
$\mu(t)$. The function $h(t)$ is determined by the inverse Laplace
transformation,
\begin{equation}\label{h_t}
h(t)=\frac{1}{2\pi i} \int_{\lambda - i\infty}^{\lambda + i\infty}
\tilde{h}(p)\exp(p\, t) dp, \quad \lambda > 0.
\end{equation}

In our following calculations, the function $h(t)$ plays the role similar
to the role of fundamental solutions (the Green functions) in the
theory of differential equations. Therefore, we call it as the
fundamental one.

It is important to emphasize that the poles $p=p_i$ of the function
$\tilde{h}(p)$ coincide with the roots $z=z_i$ of the characteristic
equation~\eqref{102} up to the multiplier $-1/T$. This means that
the fundamental solution $h(t)$ is represented as a sum of the same
exponential terms as the correlation function $C(t)$. This remark
applies to the stationarity region of parameters $\nu$ and $\mu_0$
only, where the correlation function $C(t)$ exists. In particular,
the behaviors of functions $h(t)$ and $C(t)$ at $t\rightarrow
\infty$ are the same, $h(t) \propto C(t) \propto \exp (-z_0 t/T)$.
Remind that $z_0$ is the root of Eq.~\eqref{102} with the minimal
real part.

\subsection{Solution for the case of arbitrary prehistory}

In this subsection we find the solution of the homogeneous
deterministic equation~\eqref{SIDEdet} for the general prehistory of
the process,
\begin{equation}\label{prehist}
V(t\leqslant 0)=\begin{cases}
V_<(t),&t<0,\\
V(0),&t= 0.\end{cases}
\end{equation}

The integral $\int_0^\infty dt' \mu(t')V(t-t')$ in
Eq.~\eqref{SIDEdet} can be presented as a sum of two terms,
$\int_0^t dt' \mu(t')V(t-t')$ and $\int_{-\infty}^0 dt''
\mu(t-t'')V_<(t'')$. The first one is the ordinary memory term
containing integration from the ``beginning $t'=0$ of the process
history'' to the instant moment of time $t'=t$. The second integral,
\begin{equation}\label{Z(t)}
\int_{-\infty}^0 dt'' \mu(t-t'')V_<(t'') \equiv Z(t),
\end{equation}
contains integration over the prehistory. It should be considered as
the known function $Z(t)$.

After such a representation of the integral in Eq.~\eqref{SIDEdet},
the deterministic version of the SIDE takes the form,
\begin{equation}\label{det_SIDE}
\dfrac{dV(t)}{dt} = -\nu V(t) - \int_0^{\infty} dt' \mu(t')V(t-t')
-Z(t).
\end{equation}
This equation is supplemented by the specific prehistory,
\begin{equation}\label{prehist2}
V(t\leqslant 0)=\begin{cases}
0,&t<0,\\
V(0),&t= 0.\end{cases}
\end{equation}
Now the actual prehistory $V_<(t)$ is taken into account by the
additional regular force $-Z(t)$ in Eq.~\eqref{det_SIDE}.

Applying the Laplace transformation to Eq.~\eqref{det_SIDE}, we get,

\begin{equation}\label{tildeV}
\tilde{V}(p) = \dfrac{V(0) - \tilde{Z}(p)}{p + \nu +\tilde{\mu}(p)}
= \tilde{h}(p)(V(0) - \tilde{Z}(p)).
\end{equation}
Thus, the account for the prehistory of process leads to the only
change of the fundamental solution, namely, to the appearance of
additional term $Z(p)$ in the numerator of Eq.~\eqref{tildeV}. As
expected, the expression for $\tilde{V}(p)$ contains all the poles
$p_i$ which define the fundamental solution.

\subsection{Solution for the case of exponential memory function}

In this subsection, we present in the explicit form an analytical
solution of Eq.~\eqref{SIDEdet} with the exponential memory
function, Eq.~\eqref{exp}. The Laplace image of this memory function
is
\begin{equation}
 \tilde{\mu}(p) = \dfrac{\mu_0}{T} \dfrac{1}{1+p\, T},
\end{equation}
which gives only two poles for $\tilde{h}(p)$ in
Eq.~\eqref{h_mu_Laplass}. These poles are $p_{1,2}=-z_{1,2}/T $
with
\begin{equation}\label{`MuExp_s12}
z_{1,2} = \frac{1+\nu T}{2} \pm \sqrt{\frac{(1-\nu T)^2}{4}-\mu_0}.
\end{equation}
For the sake of simplicity we consider here the prehistory
Eq.~\eqref{prehist2}. Using the inverse Laplace transformation,
Eq.~\eqref{h_t}, we find the solution,
\begin{equation}\label{MuExp_V2Exp}
\frac{V(t)}{V(0)} = A_1 \exp{(-z_1 t/T)} + A_2 \exp{(-z_2 t/T)},
\end{equation}
with
\begin{equation}\label{A}
A_1 = \frac{1-z_1}{z_2-z_1}, \quad A_2 = \frac{1-z_2}{z_1-z_2}.
\end{equation}
The analysis of poles, Eq.~\eqref{`MuExp_s12}, shows that, if the
parameters $\nu$ and $\mu_0$ satisfy the condition,
\begin{equation}\label{Parab}
\mu_0 = \frac{(1-\nu T)^2}{4},
\end{equation}
the poles $z_1$ and $z_2$ coincide, i.e., the degeneration takes
place. In this case, the solution has the form,
\begin{equation}\label{MuExp_VDeg}
V(t) = V(0)\left( 1 - \dfrac{1-\nu T}{2T}t \right) \exp{(-z t/T)},
\end{equation}
where $z = {(1+\nu T)}/{2}$. The parabola, Eq.~\eqref{Parab}, is
shown by the dashed line in Fig.\ref{ExpZones}. At $\mu_0 > (1-\nu
T)^2/4$, above the parabola, the exponential decrease of $V(t)$ is
accompanied by oscillations. These oscillations are absent below the
parabola.

Comparing Eqs.~\eqref{`MuExp_s12}, \eqref{MuExp_V2Exp} with
Eq.~\eqref{101},  one can see that the solution $V(t)$ decreases
exponentially in the same region where the \emph{random process} is
stationary and the correlation function exists. Wherein, the
asymptotic behavior of the function $V(t\rightarrow\infty)$ and
$C(t\rightarrow\infty)$ coincides. This is not surprising. Indeed,
the equations for these functions are the same, the only difference
consists in the initial conditions, see Eqs.~\eqref{nc} and
\eqref{fundam}. The memory about these conditions is asymptotically
lost at $t\rightarrow\infty$ and thus, the asymptotic solutions  for
$V(t\rightarrow\infty)$ and $C(t\rightarrow\infty)$ coincide.

In the region of parameters $\nu $ and $\mu_0 $ located to the left
of the solid lines in Fig.\ref{ExpZones}, the solution $V(t)$
exponentially increases. We are most interested in the $V(t)$
behavior on the borderlines between the stationary and
non-stationary regions. On the diffusive borderline, $\mu_0+\nu T
=0$, the pole $z_2$ in Eq.~\eqref{`MuExp_s12} vanishes, and the
solution Eq.~\eqref{MuExp_V2Exp} for $V(t)$ goes asymptotically to
the constant value $A_2$. For the oscillatory borderline, $\nu T
=-1, \, \mu_0> 1$, Eqs.~\eqref{`MuExp_s12}, \eqref{MuExp_V2Exp}, and
\eqref{A} give the harmonic solution for $V(t)$,
\begin{equation}\label{MuExp_VDeg2}
V(t) = V(0)\left[\cos (\omega t)+\frac{1}{\omega T}\sin (\omega
t)\right], \, \omega= \frac{1}{T}\sqrt {\mu_0 -1}.
\end{equation}

Another method, presented in Refs.~\cite{Kampen98,Puglisi_2009}, to
solve Eq.~\eqref{SIDEdet} with exponential memory function consists
in introducing a new auxiliary variable $U(t) = \int_{-\infty}^{t}V(t')\exp[-(t-t')/T]dt'$.  This procedure maps
the system under consideration onto a Markov process, which is
described by two ordinary differential equations. In the absence of
random force, these equations have the form,
\begin{equation}\label{puglisi}
\begin{cases}
\frac{dV(t)}{dt} = -\nu V(t) -\frac{\mu_0}{T^2}U(t), \\[6pt]
\frac{dU(t)}{dt} = V(t) -\frac{1}{T}U(t).
\end{cases}
\end{equation}
In the case of prehistory Eq.~\eqref{prehist2}, the set of equations
\eqref{puglisi} is supplemented by the initial conditions, $V(t=0) =
V(0)$ and $U(t=0) = 0$. Solving Eqs.~\eqref{puglisi} with these
initial conditions one can easily obtain results
Eq.~\eqref{`MuExp_s12} --- \eqref{MuExp_VDeg2}. Besides, the
analysis of the stability region and oscillatory/simple decay of the
correlation functions in this region,  provided in this paper for
the case of exponential memory function, is equivalent to the study
of the eigenvalues of the coupling matrix between the variables
$V(t)$ and $U(t)$.

A similar asymptotic behavior of $V(t)$ in the different regions of
$(\nu, \mu_0)$-plain takes place not only for the system with
exponential memory function but for other systems with arbitrary
$\mu(t)$ having a well-defined memory depth $T$.

\section{Movement under the action of random force}

At the beginning of this section, we show by numerical simulations
that, taking into account the random force in the Mori-Zwanzig
equation, one can observe the diffusion with memory on the lower
borderline of stationarity and the noise-induced resonance on the
upper borderline. Then we analyze the variance $ D (t) $ which
characterizes conveniently  the correlation properties of the
stochastic systems and compare the behavior of this function in
various domains in the $(\nu, \mu_0)$-plane.

\subsection{Numerical simulations}

The account of the $\sigma dW(t)$-term in Eq.~\eqref{SIDE} allows
one to describe the stochastic features of the process under
consideration. It does not change the location of stationarity
borderlines, they can still be defined by analyzing the
correspondent deterministic dynamical equation. This is the
consequence of the fact that the Gaussian noise can neither limit an
exponentially increasing solution in the non-stationarity region,
nor overcome the attraction effects in the stationarity zone.
However, the stochastic force changes the system dynamics,
especially on the stationarity borderlines.

\begin{figure}[h!]
\center\includegraphics[width=0.55\textwidth]{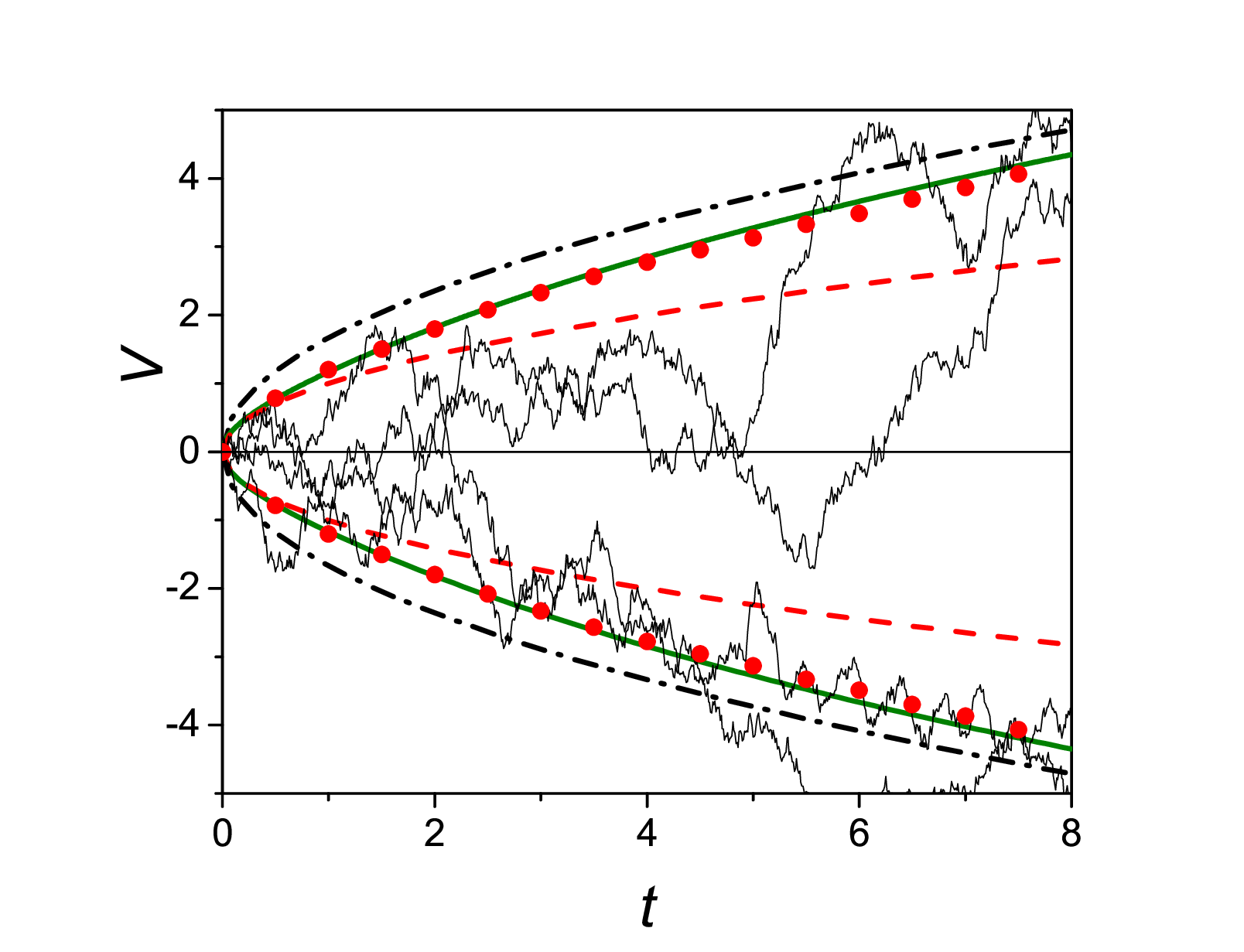} \caption{The
memory dependent diffusion for the exponential memory function and
zero prehistory $V(t\leqslant 0)=0$.  The irregular black solid
lines are the trajectories for different realizations of the
stochastic process $V(t)$ on the
diffusive borderline of stationarity. The green solid line is the
analytical result for $\pm \sqrt {D(t) }$ where $D(t)$ is the
variance, Eq.~\eqref{TPD_Diff}. The red symbols on this curve are
the results of numerical simulation obtained by the ensemble
averaging over $10^3$ realizations for each symbol. The dashed red
line presents the $\pm \sqrt {D_{\,\texttt{B}}(t)}=\pm \sigma \sqrt {t}$
dependence for the Brownian diffusion. The dash-dotted curve is the
dependence $ \pm \sigma \sqrt { t}/(1+\nu T)  $  which serves as the
asymptote for $\pm \sqrt {D(t) }$ at $t\gg T$, see
Eqs.~\eqref{TPD_Diff} and \eqref{ACB}. The parameters are: $\nu =
-0.4$, $\mu_0 = 0.4$, $T = 1$, and $\sigma = 1$.} \label{Diffus2}
\end{figure}

Irregular thin black solid lines in Fig.~\ref{Diffus2} show several
realizations of the diffusion motion for the Mori-Zwanzig equation
with exponential memory function and the zero prehistory
$V(t\leqslant 0)=0$. The parameters $\nu , T $, and $\mu_0 $ are
chosen to satisfy the condition $\nu T + \mu_0 = 0$. At first
glance, this memory-dependent diffusion does not differ from the
usual Brownian motion. However, there exists an essential difference.
To demonstrate this difference, we carried out the ensemble
averaging of $V^2(t)$ over $10^3$ realizations. The obtained
dependence  $\pm\sqrt {D(t)}=\pm \sqrt{\langle V^2(t) \rangle } $ is
plotted by the red symbols on the green solid line. In addition,
we present the similar plot for the Brownian diffusion by the red
dashed curve. The comparison of these two curves shows that the
memory dependent diffusion follows the usual Brownian motion at
small time scale $t \ll T$ only. This coincidence at short times is
not surprising. It is due to the chosen zero prehistory. However, at
$t\gtrsim T$ the memory begins to play the important role in the
diffusion. Therefore, the green solid curve in Fig.~\ref{Diffus2} deviates
from the Brownian red dashed line and tends to another asymptote with a
greater diffusion coefficient.

\begin{figure}[h!]
\center\includegraphics[width=0.5\textwidth]{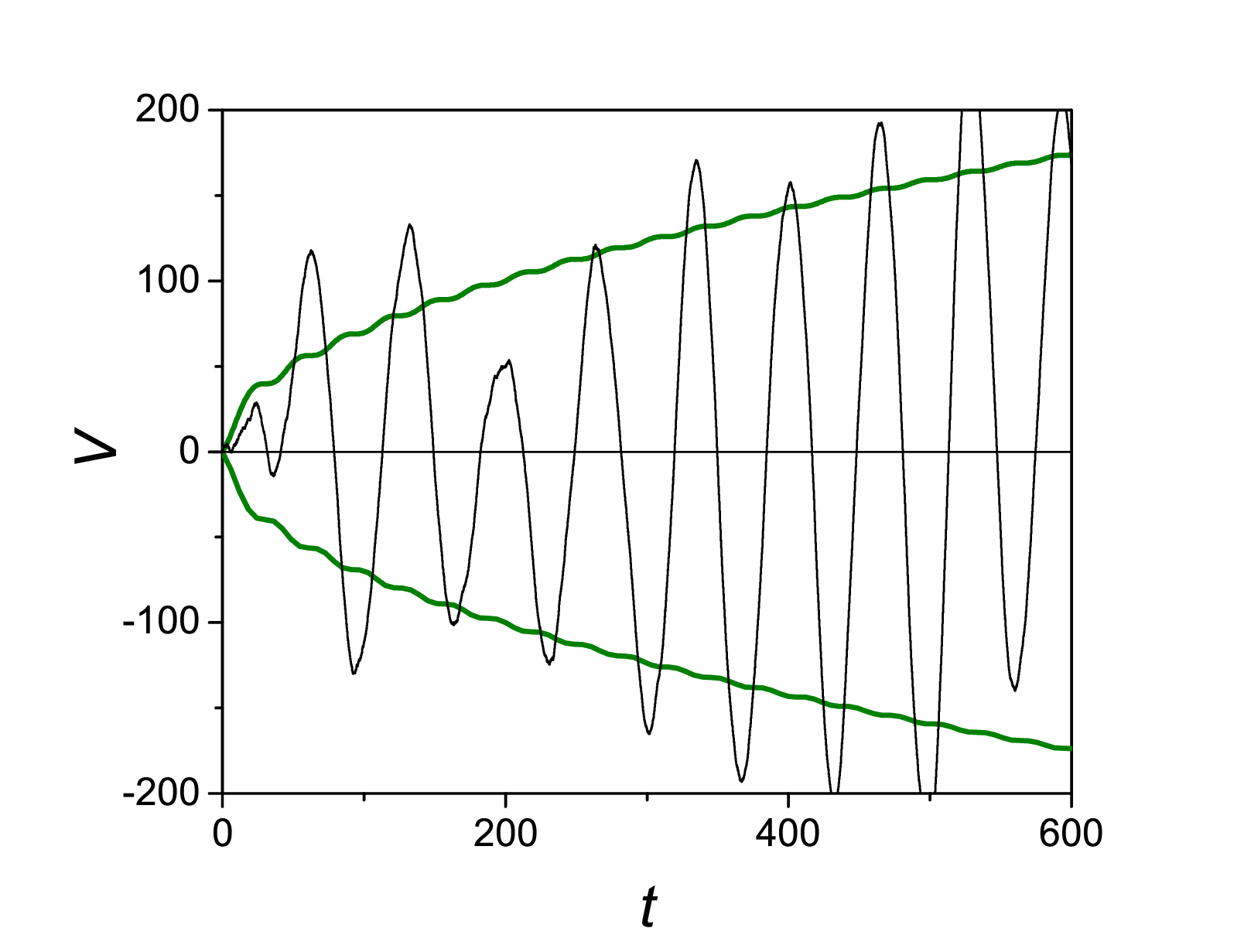}
\caption{The noise-induced resonance in the $V(t)$ process with the exponential memory function and zero prehistory. The thin
black solid line shows a realization of the stochastic process
$V(t)$ on the oscillatory borderline of stationarity. The green
solid line presents the analytical result for $\pm \sqrt {D(t)}$,
Eq.~\eqref{DTD2}. The parameters are: $\nu = -1/T$, $\mu_0 = 1.01$,
$T = 1$, and $\sigma = 1$. } \label{Stoch}
\end{figure}

Figure~\ref{Stoch} demonstrates the oscillatory motion with increasing amplitude for the Mori-Zwanzig system under the action of random force. This motion occurs with the frequency close to the frequency of self-oscillations, Eq.~\eqref{MuExp_VDeg2}. The Fourier analysis made for a 6000-length realization of the process gives an estimate $\Delta\omega / \omega \sim 0.04$ for the relative width of frequency domain of these oscillations. This means that we deal with a kind of the noise-induced resonance.

One of the most frequently discussed types of amplification of
oscillations due to external noise is the stochastic resonance (see,
e.g., Refs.~\cite{Benzi_1981,Gammaitoni,McDonnell} and references
therein). Usually, stochastic resonance is considered for the
\emph{nonlinear} systems with double-well potentials in the presence
of an external regular periodic force, and resonance occurs when the
frequency of the external force is comparable with half the
characteristic frequency of the noise-induced interwell transitions.
In the system we are considering, there are neither double-well
potentials, nor an external periodic force. In our case, the noise
does ``double duty''. The inclusion of noise leads, firstly, to the
resonant excitation of oscillations at the self-frequency $\omega$,
Eq.~\eqref{MuExp_VDeg2}. This takes place due to the presence of
frequencies in the noise spectrum, which are close to $\omega$.
Secondly, the noise leads to a subsequent increase in the amplitude
of oscillations over time. The discussed here phenomenon resembles
the well-known coherence resonance which is also observed in the
absence of an external regular periodic force~\cite{Pikovsky_1997}. However,
contrary to the coherence resonance, we consider here the
\emph{linear} systems where the noise-induced resonance occurs due
to their memory of the prehistory.

%

\subsection{Analytical study of the $V(t)$ variance}

One of the valuable characteristics of the stationary and
non-stationary random process $V(t)$ is the variance,
\begin{equation}\label{D2_def}
D(t) = \langle V^2(t)\rangle - \langle V(t)\rangle^2 .
\end{equation}
The function $D(t)$ can be easily obtained by means of the exact
solution of the Mori-Zwanzig equation~\eqref{SIDE},
\begin{equation}\label{h_V}
V(t) = V(0)h(t) + \sigma \int_0^t h(t-\tau)d W(\tau),
\end{equation}
(see, e.g., Ref.~\cite{Wang}). This formula is valid for the
specific prehistory, Eq.~\eqref{prehist2}.

Using the definition Eq.~\eqref{D2_def} and the property of the
white noise $\langle dW(t)dW(t')\rangle =\delta(t-t')dtdt'$, we
express the variance $D(t)$ in terms of the fundamental solution
$h(t)$,
\begin{eqnarray}\label{D_h}
D(t) =   \sigma^2 \int_0^t h^2(\tau) d\tau + V^2(0) [h(t)-1]^2.
\end{eqnarray}
We analyze Eq.~\eqref{D_h} considering different regions of
parameters $\nu$ and $\mu_0$, specifically, the regions of
stationarity, non-stationarity and the borderlines between them. As
far as the main properties of solutions of the Mori-Zwanzig equation
do not depend essentially on the initial value $V(0)=\langle V(t)
\rangle$, we set it to be zero, $V(0)=0$, for simplicity. We carry
out our analysis for the systems with exponential memory function.

\subsubsection{Stationarity region}

In this region, the variance Eq.~\eqref{D_h} increases with $t$ but
remains finite even at $t \rightarrow \infty$, 
\begin{equation}\label{infty_infty}
D(\infty) = \sigma^2 \int_0^\infty h^2(\tau) d\tau.
\end{equation}
Indeed, the fundamental solution $h(t)$  exponentially decreases
when increasing $t$, therefore the integral in
Eq.~\eqref{infty_infty} exists.

For the process with exponential memory function, we can carry out
an analysis of the variance $D(t)$ in more details and obtain
analytical expressions in explicit form. Substituting the function $
h (t) $ from Eq.~\eqref{MuExp_V2Exp} into Eq.~\eqref{D_h}, after
integration we get
\begin{eqnarray}\label{D2_exp}
D(t) = \sigma^2 T\!\!\!\! \sum_{\!\!i,k = 1,2}\!\!\! \dfrac{A_i
A_k}{z_i+z_k}\left\{1-
\exp\left[-(z_i+z_k)\frac{t}{T}\right]\right\} .
\end{eqnarray}
At $t\rightarrow\infty$, the exponential function in this equation
goes to zero and we obtain for $D(\infty)$,

\begin{equation} \label{D_exp_InftyInfty}
D(\infty) = \frac{1}{2}\sigma^2 T \dfrac{1 + \mu_0 + \nu T}{(\mu_0 +
\nu T)(1 + \nu T)}.
\end{equation}
As expected, the variance $D(\infty)$ diverges (tends to infinity)
if the point $(\nu, \mu_0)$ approaches the diffusive borderline (due
to the first factor in the denominator of
Eq.~\eqref{D_exp_InftyInfty}) or the oscillatory borderline (due to
the second factor in the denominator).

\subsubsection{Non-stationarity region}

In the region of non-stationarity, at least one of the roots, $z_1$
or $z_2$ in Eq.~\eqref{`MuExp_s12} has the negative real part, say
$-r$. Therefore, the main contribution to Eq.~\eqref{D2_exp} gives
the term proportional to $\exp{(2rt/T)}$. So, one should observe the
exponential increase (possibly with oscillations) of the variance at
$t \rightarrow \infty$.

\subsubsection{Solution on the diffusive borderline}

\begin{figure}[h!]
\center\includegraphics[width=0.45\textwidth]{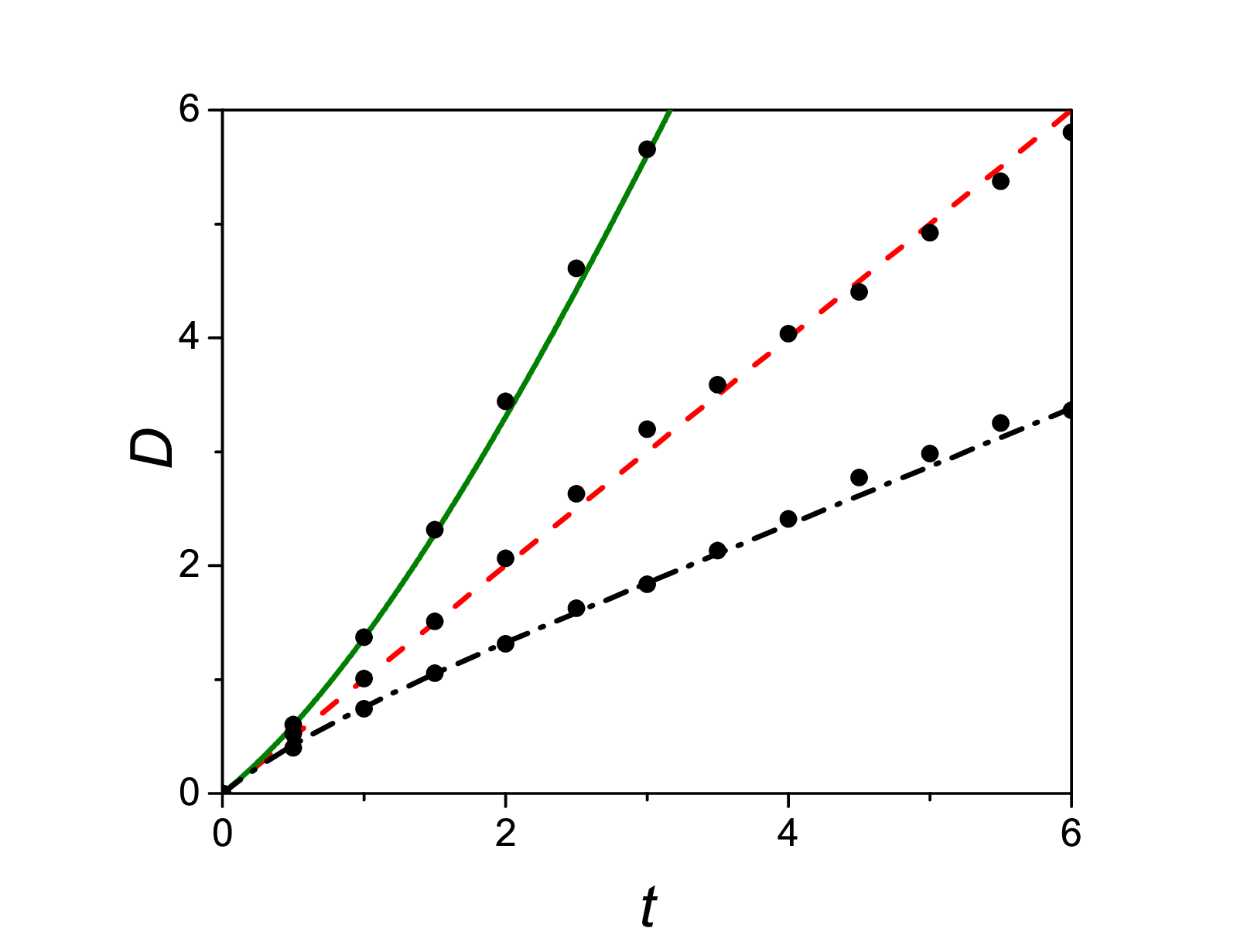} \caption{The
variance $D(t)$  on diffusive borderline for the exponential memory
function and zero prehistory at different values of $\mu_0$:
$\mu_0=0.4$ (the upper green solid curve), $\mu_0=0$ (the red
straight dashed line), and $\mu_0=-0.4$ (the lower black dash-dotted
curve). The black filled circles on these curves are the results of
numerical simulations obtained by the ensemble averaging over $10^3$
realizations for each symbol. Other parameters: $\nu=-\mu_0/T$,
$T=1$, and $\sigma=1$. } \label{DiffBord}
\end{figure}

On the line $ \nu T + \mu_0=0$, one root in Eq.~\eqref{`MuExp_s12},
say $z_1$, is real and positive, $z_1=1+\nu T=r >0$, and  the other
root is zero, $z_2=0$. Using $A_{1,2}$ in Eq.~\eqref{A}, we get the
fundamental solution,
\begin{equation}
h(t)=\frac{\nu T}{1+\nu T}\exp(-rt/T)+\frac{1}{1+\nu T},
\end{equation}
and the variance,
\begin{eqnarray}\label{TPD_Diff}
D(t)\! =\! a \,t\! + \!b \left[1-\exp(-rt/T)\right]
\!+\! c\! \left[ 1 -\exp(-2rt/T) \right]\!,
\end{eqnarray}
where
\begin{eqnarray}\label{ACB}
&&a = \frac{\sigma^2}{(1+\nu T)^2}, \quad b = \frac{2\sigma^2\nu T^2}{(1+\nu T)^3},
 \nonumber\\[6pt]
&&c = \frac{\sigma^2\nu^2 T^3}{2(1+\nu T)^3}.
\end{eqnarray}
The $D(t)$-dependence on the diffusive borderline $\nu T + \mu_0=0$
is shown in Fig.~\ref{DiffBord} for different values of $\mu_0$. One
can see that all curves follow the same straight line $D(t)=\sigma^2
t$ at $t \ll T$. This is explained by the mentioned above
circumstance: the memory does not play an essential role in the
diffusion at short time scales due to the chosen zero prehistory.
Then, at $t\gtrsim T$, the $D(t)$ curves for $\mu(t)\neq 0$ leave
the ``brownian'' asymptote $D(t)=\sigma^2 t$ and go to the other
asymptotes $D(t)=\sigma^2 t/(1+\nu T)$. In the case of positive
memory function $\mu(t)$, the curves $D(t)$ deviate upward, which
corresponds to the persistent diffusion, and for negative $\mu(t)$
the curves deviate downward, which corresponds to the
antipersistence.

\subsubsection{Solution on the noise-induced-resonance borderline}

For the exponential memory function, on the oscillatory borderline
(the vertical line in Fig.~\ref{ExpZones}, $\nu=-1/T$, $\mu_0 > 1$
), both roots, $z_1$ and $z_2=-z_1$, in Eq.~\eqref{`MuExp_s12}) are
imaginary, $z_1=ir$, $z_2=-ir$, where $r=\sqrt{\mu_0-1}$. Using the
coefficients in the fundamental solution Eq.~\eqref{A},
\begin{equation}
A_1=\frac{r +i}{2r}, \qquad A_2 = \frac{r-i}{2r},
\end{equation}
and Eqs.~\eqref{D2_exp}, \eqref{MuExp_VDeg2}, we get 

\begin{eqnarray}\label{DTD2}
&D(t)&= \frac{\sigma^2 }{2\omega ^2 T}\left[\mu_0 \frac{t}{T}+
(\mu_0-2)\frac{\sin(2\omega t)}{2 \omega T} \right. \nonumber\\[6pt]
 && \left. + 1-\cos(2\omega t)\right],\quad \omega =
 \frac{1}{T}\sqrt{\mu_0 - 1}.
\end{eqnarray}
The dependence $\pm \sqrt{D(t)}$ for the noise-induced resonance
occurring on the oscillatory borderline is shown by the green solid
line in Fig.~\ref{Stoch}. One can see that, in accordance with
Eq.~\eqref{DTD2}, the oscillations of ${D(t)}$ occur at the
frequency $2\omega$.

\section{Conclusion}

We have studied the continuous random non-markovian processes with
non-local memory and obtained new solutions of the Mori-Zwanzig
equation describing them. We have analyzed the system dynamics
depending on the amplitudes $\nu$ and $\mu_0$ of the local and
non-local memories and payed attention to the line in the ($\nu$,
$\mu_0$)-plane separating the regions with asymptotically stationary
and non-stationary behavior. We have obtained general equations for
such borderlines and considered them for three examples of the
non-local memory functions. The first example is the local, but
remote from the instant time moment $t$, memory function; the second
example is the step-wise memory function; at last, we have indicated
that Eq.~\eqref{SIDE} has an exact analytical solution for the
memory function of the exponential form.

In this paper, we have focused mainly on the system dynamics on the
borderlines of asymptotic stationarity. We have shown that there
exist two types of such borderlines with fundamentally different
system dynamics. On boundaries of the first type, corresponding to
the smaller values of $ \mu_0 $, a diffusion with memory takes
place, and on the boundaries of the second type, corresponding to
the larger values of $ \mu_0 $, the phenomenon of noise-induced
resonance occurs.

We have analyzed the dynamics of system for different prehistories
in various areas on the $ (\nu,\mu_0) $-plain in the absence of
random force. We have shown that, on the lower borderline of the
asymptotic-stationarity region, the variable $ V $ tends to a
constant value at $t\rightarrow\infty$. On the upper borderline, the
variable $ V(t\rightarrow \infty) $ goes asymptotically into
oscillatory mode with some given frequency. This means that we deal
here with the classical oscillatory motion.

Then, we have considered the system behavior under the action of
random force. We have shown that on borderlines of the first type,
corresponding to smaller values of the amplitude $\mu_0$ of
non-local memory, the diffusion with memory takes place, whereas on
borderlines of the second type, corresponding to larger values of
$\mu_0$, the phenomenon of noise-induced resonance occurs. A
distinctive feature of noise-induced resonance in the systems under
consideration is that it occurs in the absence of an external
regular periodic force. It takes place due to the presence  of
frequencies in the noise spectrum, which are close to the
self-frequency of the system.

We have analyzed also the variance of the process and compared its
behavior for regions of asymptotic stationarity and
non-stationarity, as well as for diffusive and
noise-induced-resonance borderlines between them.

The main results of this paper are valid for the processes with arbitrary memory kernel
$\mu(t)$, which is restricted by the condition $\lim_{\mathfrak{T}\rightarrow\infty} \int^\mathfrak{T} \mu(t) dt<\infty$. This means that our theory fails for the polynomial memory functions ($\mu(t)\propto t^{-\alpha}$ at $t\rightarrow\infty$ with $\alpha < 1$) for which
the integral does not converge. It would be
interesting to generalize our consideration to the non-markovian systems
with infinite memory lengths.

We have studied the memory dependent diffusion and noise-induced resonance
for the case of \emph{delta-correlated external} noise.
It seems reasonable, in future, to study the discussed phenomena for a more general case
of \emph{internal noise and long-range correlated noise} (see, e.g.,
Refs.~\cite{Wang,Goychuk1}). We believe that, in such systems,
the noise-induced resonance will not only continue to take place, but will also acquire
new interesting features.

\appendix

\section{Continuous Yule-Walker equation}

Here we present a simple derivation of Eq.~\eqref{AM_C_Phi} for the
correlation function $C(t)$ of continuous stationary process.

The exact solution Eq.~\eqref{h_V} of the Mori-Zwanzig equation
allows us to find all statistical characteristics of the system
including its correlation function. Using the definition
Eq.~\eqref{C(t)} and the property of the white noise $\langle
dW(t)dW(t')\rangle =\delta(t-t')dtdt'$, we obtain after simple
calculations the following result:
\begin{equation}\label{Cor_h}
C(t)=\lim_{t'\rightarrow\infty}C(t',t'+t)=\sigma^2 \int_0^\infty
h(\tau)h(\tau+t)d\tau.
\end{equation}
Remind that the function $h(t)$ (with the fundamental prehistory,
Eq.~\eqref{fundam}) is the solution of the deterministic version of
the Mori-Zwanzig equation,
\begin{equation}\label{h_mu_Origin}
\dot{h}(t)+\nu h(t)+\int_0^t h(t-\tau) \mu(\tau) d\tau = 0.
\end{equation}
Using the prehistory $h(t<0)=0$ of the fundamental solution, we can
replace the upper limit of integration in Eq.~\eqref{h_mu_Origin} by
$\infty$. Differentiating Eq.~\eqref{Cor_h} with respect to $t$ and
substituting $\dot{h}(\tau+t)$ from Eq.~\eqref{h_mu_Origin}, we get
the continuous analog of the Yule-Walker equation,
Eq.~\eqref{AM_C_Phi}.

\providecommand{\noopsort}[1]{}\providecommand{\singleletter}[1]{#1}%

\end{document}